 \documentclass[10pt]{aastex63}		
 \usepackage{natbib}
\def\gapprox{\lower.4ex\hbox{$\;\buildrel >\over{\scriptstyle\sim}\;$}}
\def\lapprox{\lower.4ex\hbox{$\;\buildrel <\over{\scriptstyle\sim}\;$}}

\def\af{$\alpha_F$} 
\shortauthors{Aschwanden and Scholkmann}
\shorttitle{Power Law Fits in Empirical Data}

\begin{document}
\renewcommand{\topfraction}{0.95}
\renewcommand{\bottomfraction}{0.95}
\renewcommand{\textfraction}{0.05}
\renewcommand{\floatpagefraction}{0.95}
\renewcommand{\dbltopfraction}{0.95}
\renewcommand{\dblfloatpagefraction}{0.95}


\title{Power Laws Associated with Self-Organized Criticality:
A Comparison of Empirical Data with Model Predictions} 
 
\author{Markus J. Aschwanden}
\affil{Lockheed Martin, Solar and Astrophysics Laboratory (LMSAL),
	Advanced Technology Center (ATC),
        A021S, Bldg.252, 3251 Hanover St.,
        Palo Alto, CA 94304, USA;
        e-mail: markus.josef.aschwanden@gmail.com}

\author{Felix Scholkmann}
\affil{Scholkmann Biomedical Analytics, Zurich; 
Neurophotonics and Biosignal Processing Research Group; 
Biomedical Optics Research Laboratory;
Department of Neonatology, University Hospital Zurich; 
Neuroscience Center Zurich; University of Zurich; 
ETH Zurich, Switzerland.}

\begin{abstract}
We investigate the relevance of {\sl self-organized 
criticality (SOC)} models in previously published 
empirical datasets, which includes statistical 
observations in astrophysics, geophysics, biophysics, 
sociophysics, and informatics. We study 25 
interdisciplinary phenomena with five different 
event detection and power law fitting methods. 
The total number of analyzed size distributions
amounts to 64 cases, of which 80\% are found 
to be nearly consistent ($\alpha_s=1.99\pm0.30$)
with the SOC model predictions.
The fractal-diffusive SOC model predicts power law
slopes of $\alpha_F=(9/5)=1.80$ for the flux $F$, 
$\alpha_E=(5/3)\sim1.67$ for the fluence or energy $E$, 
and $\alpha_T=2.00$ for the avalanche duration $T$.
We find that the phenomena of solar flares, earthquakes,
and forest fires are consistent with the theoretical
predictions, while the size distributions of other
phenomena are not conclusive due to neglected 
background treatment, inadequacy of power law 
fitting range, small-number statistics, and 
finite-system size effects. 
\end{abstract}

\section{	INTRODUCTION 				}  

The concept of {\sl self-organized criticality (SOC)} was 
originally proposed by Bak, Tang, and Wiesenfeld (1987) in a 
seminal article in Physical Review Letters as an explanation 
of 1/f noise and has been cited in over 10,000 publications 
since. According to the authors, SOC is a phenomenon in 
nature where “dynamical systems with extended spatial degrees of
freedom naturally evolve into self-organized critical structures 
of states which are barely stable”, leading to a power law of 
the temporal fluctuations of the system due to the combination 
of the systems’ ``dynamical minimal stability and spatial scaling''.
An interdiscipliary
approach to SOC phenomena is sketched in the book of Bak (1996).
The archetype (or SOC prototype) is the sandpile, which 
can be explored (i) by numerical simulations, as well as
(ii) by physical experiments. This dual model approach
has also been applied in other SOC phenomena, such as originally 
envisioned in the book {\sl `` How nature works''}
(Bak 1996): landscape formations,
earthquakes, solar flares, black holes, mass extinctions,
brains, river networks, traffic jams, etc. It has been 
proclaimed that {\sl power laws are the hallmarks of SOC}, 
which inspires us to study size distribution functions
(or occurrence frequency distributions) to quantify SOC 
phenomena. It was further stated that {\sl ``fractals in 
nature originate from self-organized criticality dynamical
processes''} (Bak and Chen 1989), which points us to 
fractal geometries. To clarify the nomenclature, terms
like avalanches, catastrophes, disasters, explosive
events, flare events, cataclysimic events, or epidemics, are 
interchangably used in the SOC literature.  What is 
common to all these SOC phenomena is an initial 
exponential-growth evolution in the rise phase, 
and followed by a subsequent fractal-diffusive 
decay phase. The microscopic spatial-temporal 
evolution of a SOC avalanche can be studied (i)
by computer simulations of iterative next-neighbor 
interactions in a discretized lattice grid, such 
as in the original {\sl Bak-Tang-Wiesenfeld (BTW)} model, 
or alternatively (ii) by fitting macroscopic physical 
scaling laws to observed size distributions,
such as in the novel {\sl fractal-diffusive 
self-organized criticality (FD-SOC)} model
(Aschwanden 2014, 2015; Aschwanden et al.~2013a, 2013b, 2016; 
Aschwanden and Gogus 2025).
Traditional SOC reviews can be found in Watkins et al.~(2016); 
Sharma et al.~(2016); and McAteer et al.~(2016),
or in the textbooks of Aschwanden (2011, 2025) 
and Pruessner (2012).

Although numerous power law-like size distributions have been 
found for a wide range of phenomena, there is no accepted 
physical understanding of what the numerical value of the
power law slope $(\alpha_s)$ should be (e.g., $\alpha_s=1.5$, 
1.67, 1.8 or 2.0?). The following questions remain unanswered:”
Which SOC phenomena should agree with such a theoretically
expected value $\alpha_s$? Which SOC phenomena
exhibit deviations from ideal power laws?
To what degree do background treatment,
incomplete sampling, the choice 
of fitted size ranges, and finite-size effects
affect the accuracy of power law fits?
How is the size of a SOC phenomenon defined?
How can the sizes in astrophysical, geophysical,
biophysical, sociophysical, and informatics observations
be reconciled in a self-consistent SOC theory?
Are the predicted power law slopes $\alpha_s$
dependent on the multi-farious physical mechanisms
of each SOC phenomenon? Or is there an universally
valid value that can claim to be an universal
constant? There are many more questions, but
in this study we aim to answer the questions 
raised in this paragraph. Without answering
these questions we cannot claim to understand
the physics of SOC.  

The contents of this Paper include 
a brief summary of SOC theory (Chapter 2),
observations and data analysis (Chapter 3),
discussion (Chapter 4), and
conclusions (Chapter 5).

\section{	SOC THEORY 	       }

The {\sl fractal-diffusive Self-Organized Criticality (FD-SOC)} 
model is based on four fundamental assumptions (Aschwanden 2025):
(i) The fractality, 
(ii) the scale-freeness, 
(iii) the flux-volume proportionality, and 
(iv) classical diffusion.

Instead of using the next-neighbor interactions of cellular
automata, as defined in the original SOC model of Bak (1987),
the spatial inhomogeneity of a SOC avalanche is expressed
in terms of the fractal dimension $D_d$ for the Euclidean domains
$d=1$ (lines), $d=2$ (areas), or $d=3$ (volumes).
Each fractal domain has a maximum fractal dimension of $D_d=d$,
a minimum value of $D_d=(d-1)$, and a mean value of $D_V=d-1/2$,
\begin{equation}
        D_V={(D_{\rm V,max} + D_{\rm V,min}) \over 2} = d-{1 \over 2}  \ .
\end{equation}
For most applications in the observed 3-D world, the dimensional
domain $d=3$ is appropriate, which implies a fractal dimension of 
$D_V=2.5$. The fractal volume $V$ is then defined by the standard
(Hausdorff) fractal dimension $D_V$ in 3-D and the length scale
$L$ (Mandelbrot 1977),
\begin{equation}
        V(L) \propto L^{D_V} \ .
\end{equation}
The flux is assumed to be proportional to the avalanche volume
for the case of incoherent growth ($\gamma=1$),
but can be generalized for coherent growth ($\gamma \gapprox 2$),
\begin{equation}
        F \propto V^\gamma = \left( L^{D_V} \right)^\gamma \ .
\end{equation}
The spatio-temporal evolution is approximated with the assumption 
of (classical) diffusive transport,
\begin{equation}
        L \propto T^{\beta/2} = T^{1/2} \ ,
\end{equation}
with the diffusion coefficient $\beta=1$. 

The statistics of SOC avalanches is quantified in terms of size 
distributions (or occurrence frequency
distributions) that obey the scale-free probability distribution
function (Aschwanden 2014, 2015, 2022), expressed with the power
law function, akas Zipf's law (Newman 2005),
\begin{equation}
        N(L)\ dL \propto L^{-d} dL \ .
\end{equation}
From the scale-free relationship, the power law slopes $\alpha_s$
of other SOC size parameters $s=[A,V,F,E,T]$ can be derived,
such as for the area $A$, the volume $V$, the flux $F$, 
the fluence or energy $E$, and the duration $T$. The resulting 
power law slopes $\alpha_s$ can then be obtained mathematically 
by the method of variable substitution $s(L)$, by inserting the 
inverse function $L(s)$ and its derivative $|dL/ds|$,
\begin{equation}
        N(s) ds = N[L(s)] \left| {dL \over ds} \right| dL
        = \ s^{-\alpha_s} ds \ ,
\end{equation}
such as for the flux $s=F$,
\begin{equation}
        \alpha_F = 1 + {(d-1) \over D_V \gamma} = {9 \over 5} = 1.80 \ ,
\end{equation}
for the fluence of energy $s=E$,
\begin{equation}
        \alpha_E = 1 + {(d-1) \over d \gamma} = {5 \over 3} = 1.67 \ .
\end{equation}
and for the time duration $s=T$,
\begin{equation}
        \alpha_T = 1 + (d-1)\beta/2 = 2 \ .
\end{equation}
We call this model the standard FD-SOC model, 
defined by [$d=3, \gamma=1, \beta=1$], 
while the generalized FD-SOC model allows for variable coefficients
[$d, \gamma, \beta$] and alternative dimensionalities ($d = 1, 2$).
The main prediction of the FD-SOC model that we are testing here 
is the power law slope $\alpha_F=1.80$ (Fig.~1).

\section{ 	OBSERVATIONS AND DATA ANALYSIS 		} 

The databases used in this study stem from four 
papers published over the last two decades: 
Newman (2005), Clauset et al.(2009),
and Aschwanden (2019, 2021). In addition we add
a fifth dataset (market in the column ``This Work''
in Table 1), which is equivalent to the fourth
dataset, but is processed with a different method.
These datasets yield up to five different power law
slopes $\alpha_s$ for the same SOC phenomenon, 
obtained with different event detection algorithms 
and with different power law-fitting methods, which
allows us to study systematic errors. The five
datasets contain a total of 64 cases, covering
24 empirical SOC phenomena.

\subsection{	Data Selection 		}

Each of the 64 cases listed in Table 1 refers to a
size distribution that is characterized with
a (best-fit) power law slope ($\alpha_s\pm\sigma_a$).
The simplest power law distribution function can be
quantified with one free parameter $(\alpha_s)$,
\begin{equation}
	N(s) = N_0 \ \left({s \over s_{\rm min}}\right)^{-\alpha_s} ,
\end{equation}
where $s$ is a SOC size parameter, $\alpha_s$ is the
power law slope index, $s_{\rm min}$ is the size minimum,
$N(s)$ is the occurrence frequency distribution function 
(akas size distribution), and $N_0$ is the number of 
(SOC avalanche) events per dataset.

We define a power law fitting range parameter
$q$ for each of the 64 cases,
\begin{equation}	
	 q = \log^{10} \left( { s_{\rm max} \over s_{\rm min} } \right) ,
\end{equation}
expressed in units of logarithmic decades $q$, which will be used
as a criterion to evaluate the degree of small-number statistics.

In Fig.~(1) we plot the values of the power law slopes $\alpha_s$ 
for the 64 datasets as a function of
the fitting range $q$. This scatter plot
displays some interesting properties about the uncertainties
of the individual power law slope fits. For small ranges of
$q \lapprox 3$ (left-side of diagram) we find a wide spread
of power law slope values in the range of
$\alpha_s \approx 1.5-4.3$ (marked with crosses in Fig.~1). 
In contrast, for large ranges $q \gapprox 3$ (right-side of 
diagram) we find a narrow spread of power law slopes values 
in the range of $\alpha_s = 1.99\pm0.30$ (marked with diamonds
in Fig.~1). 

Obviously, power law fits extending over $q=1$ or 2 decades 
provide unreliable fits, with large uncertainties, while
power law fits extending over $q=3$ to 6 decades can be
considered to be more reliable fits, since we expect that the
accuracy of the fits improves with larger power law-fitting 
ranges $q$, and this way overcomes small-number statistics.
Based on this bifurcation we end up with $\approx 80\%$
of acceptable fits, out of the 64 size distribution fits. 
The statistical scatter displayed in Fig.~(1) suggests
a typical error of $\sigma/\alpha_s\approx 0.3/2.0=15\%$,
provided that there a central mean value exists.
This empirical estimate of the uncertainty of power law slopes 
is about one order of magnitude larger than the published 
error estimates, typically claimed to be 
$\sigma/\alpha_s\approx 0.02/2.0=1\%$.
Thus, we conclude that the uncertainty of power law
slopes is under-estimated in most of the published 
SOC literature. The relative large uncertainties of
$\approx 15\%$ are caused by
inadequate background treatment, incomplete
sampling at the threshold of event detection,
inadequate choice of power law fitting ranges,
and finite system-size effects, as we will see
in the following discussion. The ambiguous fits
include broken (or double) power law functions,
especially for cases with cumulative distributions). 

\subsection{	Newman (2005)			}

The first dataset is from Newman (2005), 
which provides power law slopes $\alpha_s \pm \sigma$
and fitting ranges $q$ of 12 phenomena in astrophysics,
geophysics, biophysics, sociophysics, and informatics,
which can be considered as typical examples in the
overall. 
These 12 examples of power law fits are
shown in Fig.~(4) of Newman (2005), which were
obtained with the cumulative rank-order plot method,
which is defined as follows. From an event
list of a parameter $x_i, i=1, ..., n$, which is generally not sorted,
one has first to generate a rank-ordered list by ordering the events
according to increasing size,
\begin{equation}
        x_1 \le x_2 \le ... \le x_j \le ... \le x_n \ , \quad j=1, ..., n \ .
\end{equation}
The bins are generally not equidistant, except when using a
logarithmic scale, as defined by the difference between subsequent
values of the ordered $x_j$,
\begin{equation}
        \Delta x^{bin}_j = x^{bin}_{j+1} - x^{bin}_j \ .
\end{equation}
In a rank-ordered sequence of $n$ events, the probability for the largest
value is $1/n$, for events that are larger than the second-largest event it is
$2/n$, and so forth, while events larger than the smallest event occur
in this event list with a probability of unity. Thus, the cumulative
frequency distribution is simply the reversed rank order,
\begin{equation}
        N^{cum}(>x_j) = ( n+1-j ) \ , \qquad j=1,...,n \ ,
\end{equation}
and the distribution varies from $N^{cum}(>x_1)=n$ for $j=1$ to
$N^{cum}(>x_n)=1$ for $j=n$.
We can plot a cumulative frequency distribution with $N^{cum}(>x_j)$
on the y-axis versus the size $x_j$ on the x-axis. The distribution is
normalized to the number of events $n$,
\begin{equation}
        \int_{x_1}^{x_n} N(x) dx = N^{cum}(>x_1)=n \ .
\end{equation}
The uncertainty $\sigma$ of each bin is estimated to be 
\begin{equation}
	\sigma/\alpha_s={\alpha_s-1 \over \sqrt{n}} \ .
\end{equation}
This is however only a lower limit of the uncertainty,
because it includes only the statistical error
(with Poisson statistics), while
systematic errors (such as background subtraction,
inadequacy of power law fitting range, and 
finite-system effects) are not included.
Some of these systematic effects are clearly visible
in the 12 size distributions shown in Fig.~(4) of Newman (2005):
such as neglected background subtraction for the
phenomenon of solar flare peak intensity (since
there is an excess at small values of $s$);
too small fitting range for cases with $q<3$;
finite system-size effects for lunar craters,
war intensity, net worth, and name frequency.
A more realistic error can be estimated from
the spread of power law slope values as shown 
in Fig.~(1), which is of order of 
$\sigma/\alpha \gapprox 15\%$, rather than the 
statistical error of $\sigma/\alpha \approx 1\%$.

\subsection{	Clauset et al.~(2009)	}
	
The second empirical dataset is from Clauset et al.~(2009),
which doubles the cases of SOC phenomena from 12 to 
24 cases with power law slope fits $\alpha_s \pm \sigma$ and 
fitting ranges $q$ (see Table 1). Clauset et al.~(2009)
calculates the cumulative size distributions, using the 
rank-order method as it was done in Newman (2005).
The size distributions shown in Figs.~(6.1) and (6.2) in Clauset
et al.~(2009) show a number of features that constitute deviations from
ideal power law functions, such as bumps in the size distributions
(for metabolic degrees and war intensities), double (or broken) 
power laws (for earthquakes and hits to websites), and insufficient
fitting ranges of $q \approx 0.5$ to 2 
(for forest fire sizes, protein interaction degree, metabolic
degree, species per genus, birth species sightings, power blackout
customers,
financial net worth, internet degree, HTTP size, book sales,
email address books size, papers authored). Hence we find only
5 phenomena with reliable power law fits exending over
$q \ge 3$ decades (namely solar flares, earthquake
intensities, words usage, telephone calls received, and hits
to websites), which have a mean and standard deviation of
$\alpha_s=1.86\pm0.17$. 

A total of 11 phenomena were already included in the first dataset
of Newman (2005) and should have identical power law fits if they
were analyzed with identical fitting methods. However, it appears
that the choice of fitting ranges and
the {maximum likelihood estimator (MLE)} method was differently
handled in the two studies (Newman 2005; Clauset et al.~2009).
We can intercompare the results in Table 1 and find that good
agreement for most of the 11 common cases, while only three cases
exhibit significant disagreements (such as for the 
earthquake intensity, frequency of surnames, and hits to websites). 
These three cases all exhibit finite-system size effects, which
partially explain the inconsistency between the two studies
of Newman (2005) and Clauset et al.~(2009).

\subsection{	Aschwanden (2019)	}
	
A third empirical dataset (Fig.2) is published in Aschwanden (2019,
Fig.~6 therein), containing 9 cases of SOC phenomena that 
have been previously analyzed by Newman (2005) and Clauset et 
al.~(2009). In this third dataset cumulative distribution
functions were fitted for the ideal power law function
(Eq.~1),
\begin{equation}
        N_{cum}(>x) dx = \int_{x}^{x_2} n_0\ x^{-\alpha_x} dx
        = 1 + (n_{ev}-1) \left( { x_2^{1-\alpha}-x^{1-\alpha} \over
                          x_2^{1-\alpha}-{x_1}^{1-\alpha} } \right) \ .
\end{equation}
or for the {\sl thresholded (Pareto-type) power law size distribution},
\begin{equation}
        N_{cum}(>x) = \int_{x}^{x_2} n_0
        ( x + x_0 )^{-\alpha_x} dx
        = 1 + (n_{ev}-1) \left( {(x_2+x_0)^{1-\alpha_x}-(x  +x_0)^{1-\alpha_x} \over
                         (x_2+x_0)^{1-\alpha_x}-(x_1+x_0)^{1-\alpha_x} } \right) \ ,
\end{equation}
where the difference in the power law slope between the differential
size distribution function $\alpha^{diff}$ and the cumulative
size distribution function $\alpha^{cum}=\alpha^{diff}-1$ 
is already corrected.
The two size distribution functions (Eqs.~17 and 18) 
differ in the additive constant
$x_0$, which mimicks a gradual roll-over in the incompletely sampled
small-size range.

If we investigate the goodness-of-fit
$\chi^2$-values of all events in each dataset, we find 
significant deviations from ideal power laws for most of the cases.
These results differ somewhat from those obtained in Clauset et al.~(2009),
although they also find most of the data are not consistent with
an ideal power law distribution,
which is expected for fitting different functions to the observed
distributions, such as the thresholded Pareto-type power law
function used here. The evaluation
of power law deviations in Clauset et al.~(2009) is subject to
an arbitrary truncation of small events, while our method
derives a threshold automatically, which is important, because
the choice of (scale-free) inertial range boundaries affects
the outcome of the power law slope value most.
The largest deviations from ideal power laws occur
at the largest sizes, where finite-system effects dominate.

\subsection{	Aschwanden (2021)	}

The fourth dataset analyzed here is due to
Aschwanden (2021), which covers the same 9 cases as
in the previous Section (Aschwanden 2019). The fourth
method employs a 4-parameter fit that includes the
power law slope $\alpha_s$, the Pareto threshold parameter $x_0$,
the fraction of extreme events $q_{\rm pow}$, and 
the finite-system size limit $x_e$, 
\begin{equation}
        N(x) dx = n_0 \left( x_0 + x \right)^{-\alpha_x}
        \left[(1-q_{pow}) \exp{ \left( - {x \over x_e}\right) }
        + q_{pow} \right] \ dx \ .
\end{equation}
However, the differential size distribution is fitted (Eq.~10), 
rather than the
cumulative size distributions employed in the previous datasets.
Comparing the so obtained power law slopes of the two datasets
of Aschwanden (2019) and (2021), we find good agreement
for 6 out of the 9 SOC phenomena (solar flares, forest fire,
earthquake intensity, city population, word usage, frequency
of surnames), while 3 cases show significant differences
(power blackouts, terrorism, and 
website links). Two of these cases are due to small
fitting ranges, $q=1.5-1.7$ for power blackouts, and
$q=2.0-2.4$ for terrorism. The last
case (weblinks) spans over the largest fitting range 
($q=5.5$), and thus should minimize the effects of 
small-number statistics or finite-system sizes. 
It appears that the difference of the best-fit 
power-low slopes results from different weighting 
in cumulative versus differential size distributions,
in particular in the presence of large-number statistics
($q \approx 3-6$). 

\subsection{	The Fifth Method 	}

The first 3 methods deal with cumulative size distributions,
while the fourth and fifth method deal with differential
size distributions. The fourth method involves a 4-parameter
fit ($x_0, \alpha_s, q_{pow}, x_e$), but for the fifth 
method we employ a 2-parameger fit ($a_0, \alpha_s$),
in order to minimize model ambiguities. The resulting
size distributions are shown in Fig.~(4). The power law
fitting ranges $(x_1, x_2)$, where a power law function
is fitted, are demarcated with vertical
dashed lines. The systematic uncertainties are 
empirically estimated from repeating power law fits
to the log-log binned histogram $n(s)$ (of sizes $s$)
with $3 \times 3=9$  different ranges, i.e., $x_1\pm 1 {\rm\ bin}$ 
and $x_2\pm 1 {\rm\ bin}$,
yielding a mean value $\alpha_s$ and standard deviation
$\sigma$ for every case. In other words, the uncertainty
of the power law slope is estimated from the uncertainties
of the unknown fitting ranges $[x_1, x_2]$.
Another difference between the fourth and the fifth method
is the statistical weighting, which follows the
maximum likelihood estimator (MLE) method in the fourth method,
while we choose logarithmic weighting in the fifth method,
in order to give small and large events equal weighting.
It is interesting to compare the fourth
with the fifth method, because the differences of the 
the power law slopes is solely due to the parameterization
of the fitted function. Thus, the fifth method represents
also the most robust method, because it has a minimum
of two free parameters.

\section{	DISCUSSION		}

What is common to all SOC processes is the
fractality, physical scaling laws, power law-like size
distributions, and the evolution from an exponential-growth
phase to a diffusive-decay phase. 
In the following we discuss SOC properties grouped by science
disciplines (astrophysics, geophysics, biophysics, sociophysics,
informatics), as enumerated in Table 1. We critically review
the various claims of power law fits and discuss briefly some
of the physical interpretations. We acquire two different
criteria to judge the validity of the power law fits discussed
here: the mean and standard deviation $\alpha_s\pm\sigma$, and
the logarithmic power law fitting range $q$.

\subsection{	Astrophysics		}

In the astrophysics section in Table 1 we are quoting statistics 
of solar flare peak fluxes, but the universality of SOC systems
has also been identified in stellar flares (Aschwanden 
and G\"udel 2021), and in galactic, extragalactic,
and black-hole systems (Aschwanden and Gogus 2025).
What is observed in astrophysical SOC systems is
generally photon fluxes in various wavelength regimes
(gamma rays, hard X-rays, soft X-rays, extreme-ultraviolet
(EUV), ultra-violet (UV), and white-light). 
The exponential-growth phase and subsequent
fractal-diffusive phase is most conspicuosly present
in the rise and decay phase of a hard X-ray light curve.
Apparently, there is no wavelength dependence, instrumental
dependence, nor solar cycle dependence in the fitted 
power law slopes $\alpha_F$, as the data from three
different instruments (HXRBS/SMM, BATSE/CGRO, RHESSI) 
and five different power law fitting techniques
demonstrate, yielding a mean and standard deviation of 
$\alpha_F=1.790\pm0.058$, which is close to the theoretically
predicted value of $\alpha_F=1.8$.

Another astrophysical phenomenon is the impact cratering
on the moon. A SOC avalanche is created on the lunar 
surface when an asteroid originating in our solar system
hits accidentally the lunar surface, which is considered
to be a SOC process. The resulting shock wave 
propagates spherically away from the epicenter, and the 
geometric size $s$ can be measured from the size of the 
impact crater. Asteroid impacts may have occurred
over 4 billion years, during most of the lifetime 
of our solar system.
The maximum size could be as large as the lunar diameter. 
The power law slope quoted in Newman (2005), $\alpha_s
=3.14\pm0.05$, however seems not to be reliable, since
the data show a double power law that is ambiguous to fit.
 
\subsection{	Geophysics 		}

Earthquakes are among the most established SOC phenomena,
as the discovery of the Gutenberg-Richter law (Gutenberg
and Richter 1944) implies. The physical
process is believed to be caused by shifting and colliding of 
tectonic plates below the surface of the Earth. The data
quoted in Table 1 were obtained all from the same Berkeley
Earthquake Catalog, but we find different power law slopes:
$\alpha_s=3.04\pm0.04$ (Newman 2005),
$\alpha_s=1.64\pm0.04$ (Clauset et al.~2009),
$\alpha_s=1.85\pm0.01$ (Aschwanden 2019), 
$\alpha_s=1.71\pm0.01$ (Aschwanden 2021), and
$\alpha_s=2.28\pm0.08$ (This work).
The first method indicates an extreme outlier, 
which is most likely a misprint in Table 1
of Newman (2005). The second method indicates 
an invalid fit in Fig.~6.2 of Clauset et al.~(2009).
The third and fourth method are more consistent
with each other and are fitted over a large
range of $q=4.3-4.5$ decades. Averaging the
latter two methods we obtain a more reliable
result of $\alpha_s=1.78\pm0.07$, which is close
to the predicted FD-SOC value of $\alpha_s=1.80$.

Forest fires (akas wild fires) are very common 
in California, which ironically is also true for 
earthquakes. The basic mechanism of a spreading
forest fire is based on next-neighbor 
interactions in form of flammable fuel that is
contained in dry conditions in the woods and bushes. 
Well-known trigger mechanisms are sparks from
power lines in the presence of strong winds,
agricultural grass burning, uncontrolled combustible 
vegetation, or arson. In our comparison
we find three different values:
The first fit ($\alpha_s=2.2\pm0.3$, Clauset et al.~2009),
shows a broken power law shape, which introduces 
some ambiguity in the choice of the fitted function.
The second fit 
($\alpha_s=1.53\pm0.01$, Aschwanden 2019) 
exhibits a prominent finite-system size effect, 
while the third fit 
($\alpha_s=1.77\pm0.01$, Aschwanden 2019)  
represents the best fit, with a large fitting range
of $q=3.8$.
 
\subsection{	Biophysics 		}

Four examples of SOC processes in biophysics
have been proposed (Clauset et al.~2009): 
(i) The degrees (i.e., numbers of distinct 
interaction partners) of proteins in the 
partially known protein-interaction network 
of the yeast Saccharomyces cerevisiae;
(ii) The degrees of metabolites in the
metabolic network of the bacterium
Escherichia coli; 
(iii) The number of species per genus
of mammals, including also recently
extinct species; and
(iv) The number of sightings of birds of
different species in the North American
Breeding Bird Survey for 2003.
These four examples, however, show power laws
only in the smallest fitting ranges of
$q=1.0$ to 1.5 (Table 1), and 
thus are questionable candidates for power law
size distributions.

The scale-free topology of protein interaction 
and metabolic networks (i.e., some proteins
(nodes) have many more interactions (edges) 
than others) and their balance between 
robustness (i.e. ability to withstand perturbations) 
and fragility (sensitivity to small changes)
are characteristics of SOC. In terms of the 
distribution of the number of species, ecological
systems could exhibit SOC due to the dynamics of 
interactions (competition, cooperation)
between species and the environment, as well as 
between species. Sightings of species in
the wild (such as birds) also appear to be the 
result of complex self-organized interactions of
members of that species with each other and with 
the environment, leading to SOC.

\subsection{	Sociophysics 		}

Sociophysics (or social physics) is a novel
science discipline that uses mathematical
tools or physical models to understand the
behavior of human groups or crowds.   
A human group can be defined by a
common characteristic, such as citizenship,
power blackout customers, soldiers in wars, 
terrorist attacks,
religions, financial net worth of wealthy
people, or stock market fluctuations (Table 1). 
Each of these characteristis can be quantified
by the size or number of a particular human group.
In a wider context, the evolution of a
SOC process can be associated with a single event,
which displays an instability or 
exponential-growth phase (once
an event is triggered above some threshold),
and a subsequent diffusive decay process.

For example, the global growth of populations
is approximately exponential in a society
that has a growth rate of at least two 
humans per lifetime. Since the population
multiplies due to next-neighbor 
(or other local) interactions, 
the statistics of city
sizes reflects the number of their 
populations, for which we then expect
a power law distribution with a slope of
$\alpha_s=1.80$,
according to the FD-SOC model. A value close to this
expectation is found at $\alpha_s=1.79\pm0.01$
(Aschwanden 2019), but all five methods listed
in Table 1 for population of cities exhibit
distinct deviations from ideal power laws,
or narrow fitting ranges. This indicates a need
for a modified SOC model for the case of city
populations (such as Eqs.~10, 11). 
Differences in the power law slope size distributions
have been noticed for different countries,
regions of wars, genocides, famines, 
epicemics, pandemics, and climates.

\subsection{	Informatics (Literature)		}

SOC phenomena have been discovered in literature as well as
in global networks. Both applications require
computer-based software. In the two studies
of Newman (2005) and Clauset et al.~(2009)
we find the following literature-based statistics:
words usage (the frequency of occurrence of unique
words in the novel {\sl Moby Dick} by Herman Melville),
book sale statistics (number of best-sellers sold in
the US during 1895-1965),
frequency of surnames (the frequencies of occurrence
of USA family names in the 1990 USA census),
citation of papers (the number of citations of
scientific papers published in 1987),
and papers authored or co-authored by mathematians.
The sizes of these literature-based SOC phenomena
is measured by the number of words, book titles, surnames,
and references in journal publications. The most accurate 
results are obtained for word usages, based on the 
availability of large fitting ranges ($q=3.5-4.0$),
based on five different methods (Table 1). The mean value
of these five different methods is $\alpha_s=2.07\pm0.12$,
which is only slightly (but significantly) higher than 
the theoretically predicted FD-SOC model,
with $\alpha=1.80$, and thus may require a modified
(generalized) SOC model.

\subsection{	Informatics (Global Networks)		}

The last group of putative SOC phenomena in the discipline
of informatics deals with global networks. It contains:
telephone calls received (by customers of AT\&T's long
distance telephone service in the USA during a single day),
internet degree (groups of computers under single 
administrative control on the internet),
HTTP size (the number of bytes of data received in
response to HTTP web requests from computers at a
large research laboratory), 
email address books sizes (at an
university, in units of kilobytes),
hits to websites (from AOL users), and
links to websites (the numbers of hyperlinks
to websites). 

The most reliable size distribution
of this group includes telephone calls, with 
a power law range of $q=4.0$ decades,
which reveals an averaged power law slope of $2.16\pm0.07$.
Links to websites are also found
to cover a large power law range of $q=5.5$, 
but exhibit quite different values, such as
$2.34\pm0.01$ for the cumulative size distribution,
$1.49\pm0.02$ and $1.33\pm0.02$ 
for the differential size distribution.
This is an unexpected behavior. 
It demonstrates that differential size distribution
fits are more accurate than cumulative size 
distributions, even for large fitting ranges
($q \gapprox 5$). 

Also we note that website links
have a power law slope that is not close
to the theoretical prediction, namely
$\alpha_s=1.49\pm0.02$ and 
$\alpha_s=1.37\pm0.01$. 
This implies that these observations cannot
be explained with the standard FD-SOC model.
One possibility is that website links are
subject to different dimensionalities,
such as 2-D topologies ($d=2$), rather than
to 3-D topologies ($d=3$). If this is the case,
the prediction of the (generalized) FD-SOC) is
$\alpha_E=1.50$ (Eq.~8 with d=2), which is 
closer to the observations, ($1.49\pm0.02$;
$1.37\pm0.01$). The physical model would have
to do with the topology of electronics 
in a central processing unit (CPU) and 
computer architecture in general.

\section{	CONCLUSIONS	}

We turn now back to the questions raised in the
Introduction and formulate the answers based
on the conclusions made from this study. 

\begin{enumerate}
\item{\sl Do we have a physical understanding 
what the power law slope value $\alpha_s$ 
of SOC avalanches should be?
\rm There is virtually no existing theory that predicts 
quantatively the power law slope $\alpha_s$ 
of the size distribution of SOC avalanches,
except for the {\sl fractal-diffusive (FD-SOC)} model
(Section 2). The physical understanding is based
on the four fundamental assumptions of fractality,
scale-freeness, flux-volume proportionality in
incoherent fluxes, and classical diffusion, which
all are universal by themselves too.}

\item{\sl What is the power law value and which SOC 
phenomena should agree with such a theoretically 
expected value $\alpha_s$?
\rm The theoretical (FD-SOC) model predicts
power law like size distributions with power law
slopes of 
$\alpha_F=(9/5)=1.80$ for the flux $F$,
$\alpha_E=(5/3)\sim1.67$ for the fluence or energy $E$, and
$\alpha_T=2.00$ for the avalanche duration $T$.
These predictions should be applicable to most of the
SOC phenomena with power law-like size distributions,
especially for datasets with complete sampling over
some size range, large number statistics ($n \gapprox 10^4$), 
and large fitting size ranges ($q \gapprox 3$).}

\item{\sl Which SOC phenomena exhibit deviations 
from ideal power laws? 
\rm The size distributions of
SOC phenomena with incomplete sampling near the
threshold, with small-number statistics 
($n \lapprox 10^2$), with small fitting ranges
($q \lapprox 2$), and with finite-system size 
effects are likely to exhibit significant
deviations from ideal (straight) power law size
distribution functions. Although finite-system
size effects are modeled with an exponential
fall-off at the largest sizes, the true 
distribution function could deviate from an
exponential fall-off.}

\item{\sl To what degree do background treatment,
incomplete sampling, the choice
of fitted size ranges, and finite-size effects
affect the accuracy of power law fits?
\rm Neglect of background subtraction of SOC-unrelated
fluxes produces an excess on top of the size
distribution for small event sizes, and thus
steepens the power law slopes at the small events.
A small size range and/or small-number statistics
produces a curved function for the size distributions
that cannot be fitted with a straight power law function
(see Fig.~2).
Finite-system size effects cause a steepening of the
slope for large size events. All these effects
enlarge the uncertainty of power law fits.}

\item{\sl How is the size of a SOC phenomenon defined?
\rm The size of a SOC phenomenon is implicitly defined
by the magnitude of an event, which can be the peak flux
of electro-magnetic (incoherent) emission in the case
of solar, stellar or  astrophysical sources, the
magnitude of an earthquake, the burned area of a
forest fire, fluctuations in a metabolic network,
the number of humans in city populations, 
wars, terrorist attacks, religions, or financial groups.
For the discipline of informatics, the sizes of SOC
phenomena may be defined by the number of words, books,
surnames, publications, telephone calls, or other
computer-based statistics occurring in global networks.}

\item{\sl How can the sizes in astrophysical, geophysical,
biophysical, sociophysical, and informatics observations
be reconciled in a self-consistent SOC theory?
\rm The unifying property of SOC avalanches is their
common nonlinear evolution, which consists of an initial
exponential-growth phase, followed by a fractal-diffusive 
decay phase, triggered by an instability after the SOC
avalanche exceeds a global threshold. This brings us back
to the sandpile paradigm, which is a mechanical analog,
but instabilities can occur also in electro-magnetic emission,
thermodynamics, hydrodynamics, magnetohydrodynamics (MHD), 
high temperature plasmas, as well as in human-created settings, 
such as literature statistics and computer-based electronics.}

\item{\sl Are the predicted power law slopes $\alpha_s$
dependent on the multi-farious physical mechanisms
of each SOC phenomenon?
\rm No! In order to understand various SOC systems, 
we do not have to worry about magnetic reconnection
in solar flares, tectonic plate motions in earthquakes,
or the influence of global warming in Californian forest fires.
The self-consistency of a single power law slope value
($\alpha_F \approx 1.80$) in size distributions of fluxes
can be considered to be established for the phenomena
of solar flares, earthquakes, and forest fires, 
according to the statistics of datasets with selected
(reliable) power law fits. All other phenomena (listed in
Table 1) are possible candidates for SOC events, but
cannot be confirmed here because of unreliable slopes
caused by insufficient ranges, small-number statistics, 
and unknown finite system-size effects.}

\item{\sl Is there an universally valid value that can 
claim to be an universal constant?
\rm Based on the self-consistency of SOC avalanche
fluxes, fluences or energies, and durations in different
SOC phenomena, we can consider their power law slopes,
$\alpha_F=(9/5)$, $\alpha_E=(5/3)$, $\alpha_T=2$,
as universal constants, because they do not vary 
for different SOC phenomena.}

\item{\sl Is the original BTW model obsolete?
\rm Self-organized criticality models have been
developed into two directions: the BTW model that
is based on microscopic next-neighbor interactions,
while the FD-SOC model replaces the microscopic
structure with the macroscopic model of physical
scaling laws in terms of fractal diffusion.
Since the BTW model represents a phenomenological
approach without any physics-based explanation,
the FD-SOC model is preferable as a physical
framework.} 
\end{enumerate}

For future work we suggest power law fitting with 
multiple databases, multiple instruments, and
large-number statistics, in order to study
systematic errors and deviations from true power laws.
Theoretical modeling of SOC phenomena may also include
information on the spatio-temporal nonlinear evolution.
Once we characterized SOC parameters with sufficient
statistics we may earn the benefit of reliable
(statistical) predictions of extreme events. 

\acknowledgments
{\sl Acknowledgements:}
This work was stimulated by the organizers of a workshop on
{\sl ``Mechanisms for extreme event generation'' (MEEG)} at the
Lorentz Center at Snellius, Leiden, The Netherlands, July 8-12, 2019,
organized by Drs. Norma Bock Crosby, Bertrand Groslambert,
Alexander Milovanov, Jens Juul Rasmussen, and Didier Sornette.
The author acknowledges the hospitality and partial support of two
previous
workshops on ``Self-Organized Criticality and Turbulence'' at the
{\sl International Space Science Institute (ISSI)} at Bern, Switzerland,
during October 15-19, 2012, and September 16-20, 2013, as well as
constructive and stimulating discussions with
Sandra Chapman,
Paul Charbonneau,
Aaron Clauset,
Norma Crosby,
Michaila Dimitropoulou,
Manolis Georgoulis,
Stefan Hergarten,
Henrik Jeldtoft Jensen,
James McAteer,
Shin Mineshige,
Laura Morales,
Mark Newman,
Naoto Nishizuka,
Gunnar Pruessner,
John Rundle,
Carolus Schrijver,
Surja Sharma,
Antoine Strugarek,
Vadim Uritsky,
and Nick Watkins.
This work was partially supported by NASA contracts NNX11A099G
``Self-organized criticality in solar physics'',
NNG04EA00C of the SDO/AIA instrument,
and NNG09FA40C of the IRIS instrument.

\clearpage

\section*{      References      }
\def\ref#1{\par\noindent\hangindent1cm {#1}}

\ref{Aschwanden, M.J. 2011,
        {\sl Self-Organized Criticality in Astrophysics. The Statistics
        of Nonlinear Processes in the Universe},
        Springer-Praxis: New York, 416p.}
\ref{Aschwanden, M.J. 2013a, in {\sl Theoretical Models of SOC Systems},
        chapter 2 in {\sl Self-Organized Criticality Systems}
        (ed. Aschwanden M.J.), Open Academic Press: Berlin, Warsaw, p.21.}
\ref{Aschwanden,M.J. 2013b,
        {\sl Self-Organized Criticality Systems in Astrophysics (Chapter 13)},
        in "Self-Organized Criticality Systems" (ed. Aschwanden,M.J.),
        Open Academic Press: Berlin, Warsaw, p.439.}
\ref{Aschwanden, M.J. 2014,
        {\sl A macroscopic description of self-organized systems and
        astrophysical applications}, ApJ 782, 54.}
\ref{Aschwanden, M.J. 2015,
        {\sl Thresholded power law size distributions of instabilities
        in astrophysics}, ApJ 814.}
\ref{Aschwanden, M.J., Crosby, N., Dimitropoulou, M., Georgoulis, M.K.,
        Hergarten, S., McAteer, J., Milovanov, A., Mineshige, S.,
        Morales, L., Nishizuka, N., Pruessner, G., Sanchez, R.,
        Sharma, S., Strugarek, A., and Uritsky, V. 2016,
        {\sl 25 Years of Self-Organized Criticality: Solar and
        Astrophysics}, Space Science Reviews 198, 47.}
\ref{Aschwanden, M.J. 2019, {\sl Self-organized criticality in
	solar and stellar flares: Are extreme events scale-free?},
	ApJ 880:105 (16pp.)}
\ref{Aschwanden, M.J. 2021,
        {\sl Finite system-size effects in self-organizing criticality 
	systems}, ApJ 909, 69.}
\ref{Aschwanden, M.J. and Guedel, M. 2021, {\sl Self-organized
	criticality in sellar flares}, ApJ 910, id.41, 16pp.}
\ref{Aschwanden, M.J. 2022,
        {\sl The fractality and size distributions of astrophysical
        self-organized criticality systems},
        ApJ 934 33}
\ref{Aschwanden, M.J. and Gogus, E. 2025,
	{\sl Testing the universality of self-organized criticality
	in galactic, extra-galactic, and black-hole systems},
	ApJ, 978:19 (11pp).}
\ref{Aschwanden, M.J. 2025,
	{\sl Power Laws in Astrophysics. Self-Organzed Criticality
	Systems}, Cambridge University Press: Cambridge.}
\ref{Bak, P., Tang, C., and Wiesenfeld, K. 1987,
        {\sl Self-organized criticality: An explanation of 1/f noise},
        Physical Review Lett. 59(27), 381.}
\ref{Bak, P. and Chen, K. 1989,
        {\sl The physics of fractals},
        Physica D  38, 5-12.}
\ref{Bak, P. 1996,
        {\sl How Nature Works. The Science of Self-Organized Criticality},
        Copernicus: New York.}
\ref{Clauset, A., Shalizi, C.R., and Newman, M.E.J. 2009, 
        {\sl Power-law distributions in empirical data},
        SIAM Review 51/4, 661-703.}
\ref{Gutenberg, B. and Richter, C.F. 1944,
	{\sl Frequency of earthquakes in California},
	Bull.Seismological Society of America, Vol.34, issue 4, pp. 185-188.}
\ref{Mandelbrot, B.B. 1977,
        {\sl Fractals: form, chance, and dimension}, Translation of
        {\sl Les objects fractals}, W.H. Freeman, San Francisco.}
\ref{McAteer,R.T.J., Aschwanden,M.J., Dimitropoulou,M., Georgoulis,M.K.,
        Pruessner, G., Morales, L., Ireland, J., and Abramenko,V. 2016,
        {\sl 25 Years of Self-Organized Criticality: Numerical Detection
        Methods}, SSRv 198, 217.}
\ref{Newman,M.E.J. 2005,
        {\sl Power laws, Pareto distributions and Zipf's law},
	Contemporary Physics, Vol. 46, issue 5, pp.323-351.}
\ref{Pruessner, G. 2012, {\sl Self-Organised Criticality. Theory, Models
        and Characterisation}, Cambridge University Press: Cambridge.}
\ref{Sharma,A.S., Aschwanden,M.J., Crosby,N.B., Klimas,A.J., Milovanov,A.V.,
        Morales,L., Sanchez,R., and Uritsky,V. 2016,
        {\sl 25 Years of Self-Organized Criticality: Space and Laboratory
        Plamsas}, SSRv 198, 167.}
\ref{Watkins, N.W., Pruessner, G., Chapman, S.C., Crosby, N.B., and Jensen, H.J.
        {\sl 25 Years of Self-organized Criticality: Concepts and 
	Controversies}, 2016, SSRv 198, 3.}

\clearpage

\begin{table}
\begin{center}
\caption{Power law slope $a_F$ fits and ranges $q$ of 
emprical datasets from 5 different publications 
(Newman 2005, Clauset et al.~2009, Aschwanden 2019,
2021, and this work).}
\normalsize
\medskip
\begin{tabular}{l|ll|ll|ll|ll|ll|}
\hline

             Phenomenon         & Newman       &     & Clauset       &     & Aschwanden    &     & Aschwanden    &     & This &     \\
                      	        & (2005)       &     & et.al.(2009)  &     & (2019)        &     & (2021)        &     & Work &     \\
                                & \af          & $q$ & \af           & $q$ & \af           & $q$ & \af           & $q$ & \af  & $q$ \\
\hline
\hline
{\underbar{\bf Astrophysics}:}  &               &     &               &     &               &     &               &    &      &     \\
solar flares HXRBS             & 1.83$\pm$0.02 & 3.0 & 1.79$\pm$0.02 & 3.5 & 1.70$\pm$0.02 & 3.8 & 1.75$\pm$0.02 & 4.0 &      &     \\
solar flares BATSE             &               &     &               &     & 1.81$\pm$0.02 & 3.3 &               &     &      &     \\
solar flares RHESSI            &               &     &               &     & 1.86$\pm$0.02 & 3.7 &               &     &      &     \\
lunar craters                  & 3.14$\pm$0.05 & 2.1 &               &     &               &     &               &     &      &     \\ 
\hline
{\underbar{\bf Geophysics:}}	&		&     &	              &	    &               &     &               &      &    &     \\
earthquake intensity           & 3.04$\pm$0.04 & 2.5 & 1.64$\pm$0.04 & 3.0  & 1.85$\pm$0.01 & 4.5  & 1.71$\pm$0.01 & 4.3 & 2.28$\pm$0.08 & 3.2 \\
forest fire size               &               &     & 2.20$\pm$0.30 & 0.5 & 1.53$\pm$0.01 & 4.0  & 1.77$\pm$0.01 & 3.0  & 1.76$\pm$0.21 & 3.8 \\
\hline
{\underbar{\bf Biophysics:}}   &		&     & 	      &	    &               &     &               &      & & \\
protein interaction degres     &               &     & 3.10$\pm$0.30 & 1.5 &               &     &               &      & & \\
metabolic degree 	       &               &     & 2.80$\pm$0.10 & 1.0  &               &     &               &     & & \\
species per genus 	       &               &     & 2.40$\pm$0.20 &	1.5 &               &	  &               &      & & \\
bird species sightings         &               &     & 2.10$\pm$0.20 & 1.0 &               &     &               &       & & \\
\hline
{\underbar{\bf Sociophysics:}} &		&	        &	        &               &                         & & \\
population of cities           & 2.30$\pm$0.05 & 2.0 & 2.37$\pm$0.08 & 2.0 & 1.79$\pm$0.01 & 2.0 & 1.59$\pm$0.01 & 4.0    & 1.96$\pm$0.14 & 4.4 \\
power blackout customers       &               &     & 2.30$\pm$0.30 & 1.5 & 1.90$\pm$0.13 & 1.5 & 1.35$\pm$0.09 & 1.7    & 1.78$\pm$0.18 & 2.4 \\
intensity of wars 	       & 1.80$\pm$0.09 & 2.0 & 1.70$\pm$0.20 & 2.0 &               &     &               &        &               &     \\
terrorist attack severity      &               &     & 2.40$\pm$0.20 & 2.0 & 2.28$\pm$0.04 & 2.0 & 2.64$\pm$0.04 & 2.4    & 2.20$\pm$0.16 & 3.0 \\ 
religious followers            &               &     & 1.80$\pm$0.10 & 2.0 &               &     &               &        & & \\
financial net worth            & 2.09$\pm$0.04 & 2.0 & 2.30$\pm$0.10 & 1.5 &               &     &               &        & & \\ 
\hline
{\underbar{\bf Informatics}} {\underbar{\bf Literature:}} &	&     & 	      &	    &               &     &       & & \\
words usage 	                & 2.20$\pm$0.01 & 4.0 & 1.95$\pm$0.02 & 4.0 & 1.99$\pm$0.02 & 3.5 & 2.12$\pm$0.02 & 3.6   & 1.83$\pm$0.12 & 3.7\\
sales of books                 & 3.51$\pm$0.16 & 0.5  & 3.70$\pm$0.30 & 1.0 &               &     &               &       & & \\
frequency of surnames          & 1.94$\pm$0.01 & 1.5 & 2.50$\pm$0.20 & 2.0 & 2.50$\pm$0.05 & 2.0 & 2.58$\pm$0.05 & 2.2    & 2.07$\pm$0.05 & 2.3\\
citation to papers             & 3.04$\pm$0.02 & 1.5 & 3.16$\pm$0.06 & 2.0 &               &     &               &        & & \\ 
papers authored                &               &     & 4.30$\pm$0.10 & 1.0 &               &     &               &        & & \\
\hline
{\underbar{\bf Informatics}} {\underbar{\bf Networks:}}&  &     & 	      &	    &               &     &       & & \\
telephone calls received 	& 2.22$\pm$0.01 & 3.0  & 2.09$\pm$0.01 & 4.0 &               &	  &               &       & & \\
internet degree 	        &               &     & 2.12$\pm$0.09 & 2.5 &               &	  &               &       & & \\
HTTP size (kilobytes)          &               &     & 2.48$\pm$0.05 & 2.5 &               &     &               &        & & \\
email address books size       &               &     & 3.50$\pm$0.60 & 1.0 &               &     &               &        & & \\
hit to websites               & 2.40$\pm$0.01 & 4.2  & 1.81$\pm$0.08 & 3.0  &               &     &               &      & & \\
link to websites              &               &     & 2.34$\pm$0.01 & 2.0 & 2.34$\pm$0.00 & 5.5  & 1.49$\pm$0.02 & 5.5   &  1.33$\pm$0.02 & 8.0\\
\hline
\end{tabular}
\end{center}
\end{table}
\clearpage

\begin{figure}
\centerline{\includegraphics[width=1.0\textwidth]{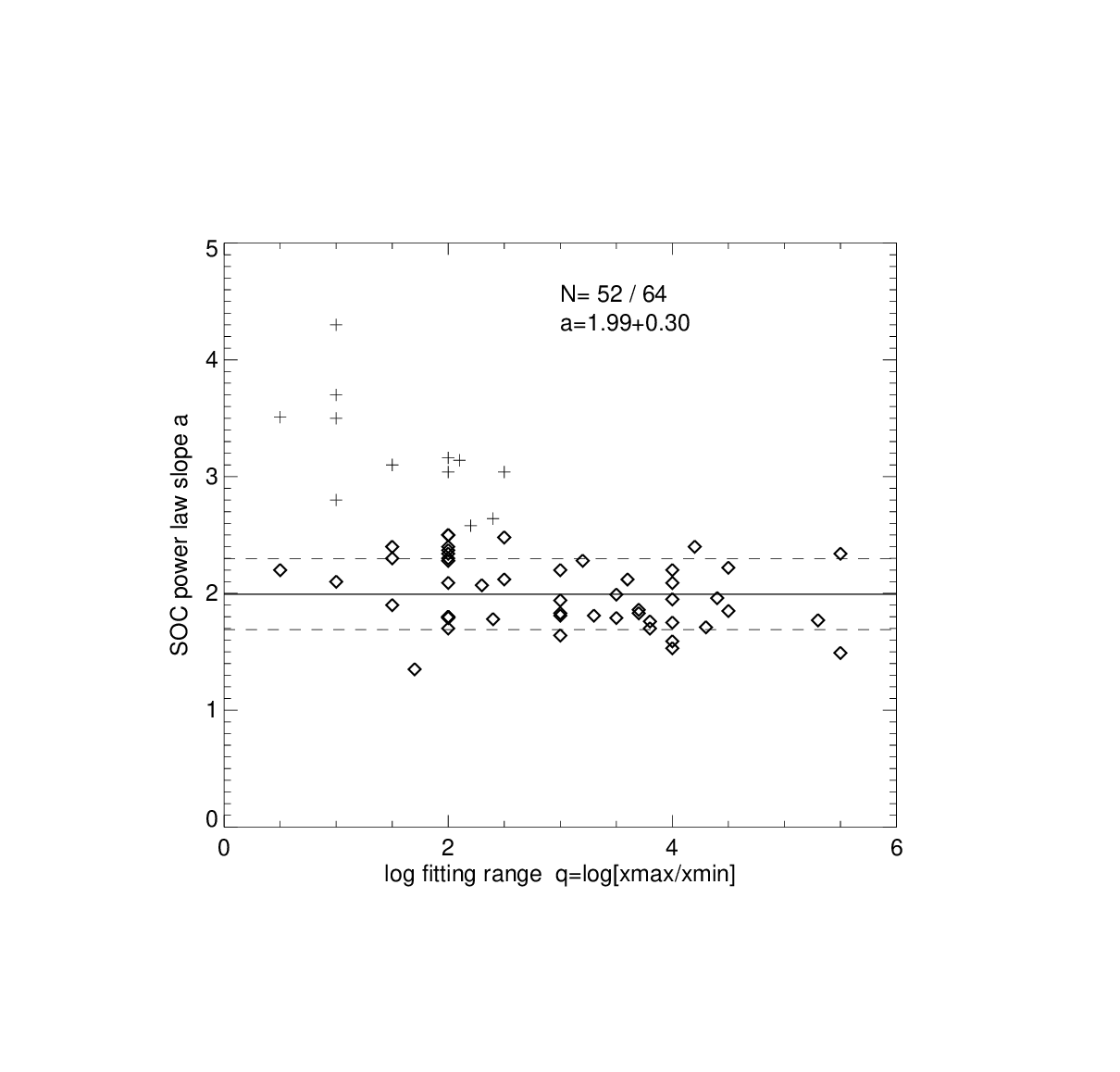}}
\caption{Best-fit power law slopes $\alpha_s$ 
of 64 different published datasets, as function of the 
fitting range $q=\log (x_{\rm max}/x_{\rm min})$.
The dataset can be subdivided into two groups: 
12 cases of outliers with $\alpha > 2.5$ (crosses), and
52 cases with $\alpha \le 2.5$ (diamonds). The latter
group has a mean and standard deviation of
$\alpha=1.99\pm0.30$.}
\end{figure}
\clearpage

\begin{figure}
\centerline{\includegraphics[width=1.0\textwidth]{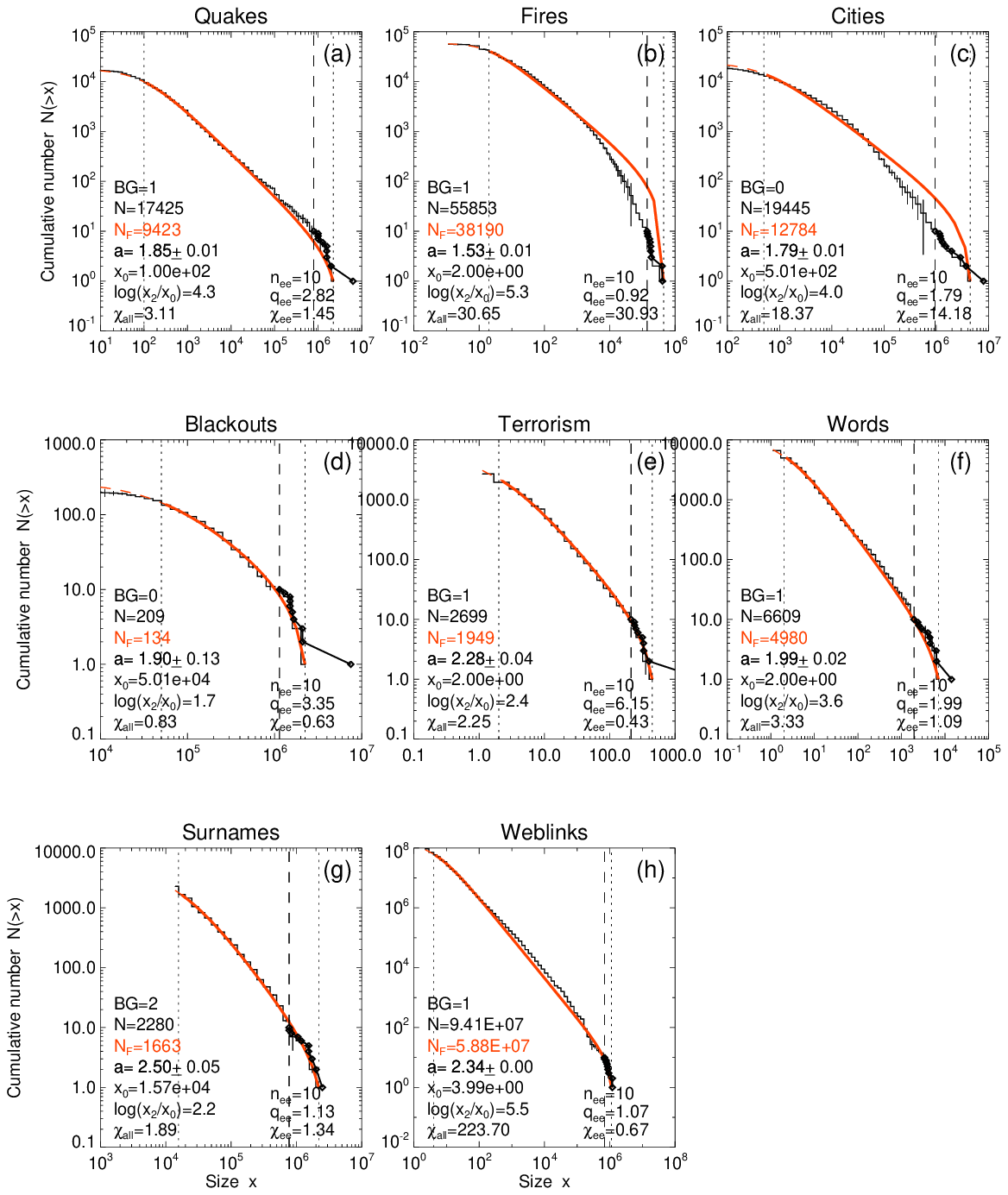}}
\caption{Cumulativek power law size distributions are shown
for eight empirical datasets from Clauset et al.~(2019):
earthquakes (a), forest fires (b), population of cities (c),
electric power blackouts (d), terrorist attack severity (e),
count of words (f), frequency of surnames (g), and links
to websites (h), [adapted from Figure 6 in Aschwanden 2019].}
\end{figure}

\begin{figure}
\centerline{\includegraphics[width=1.0\textwidth]{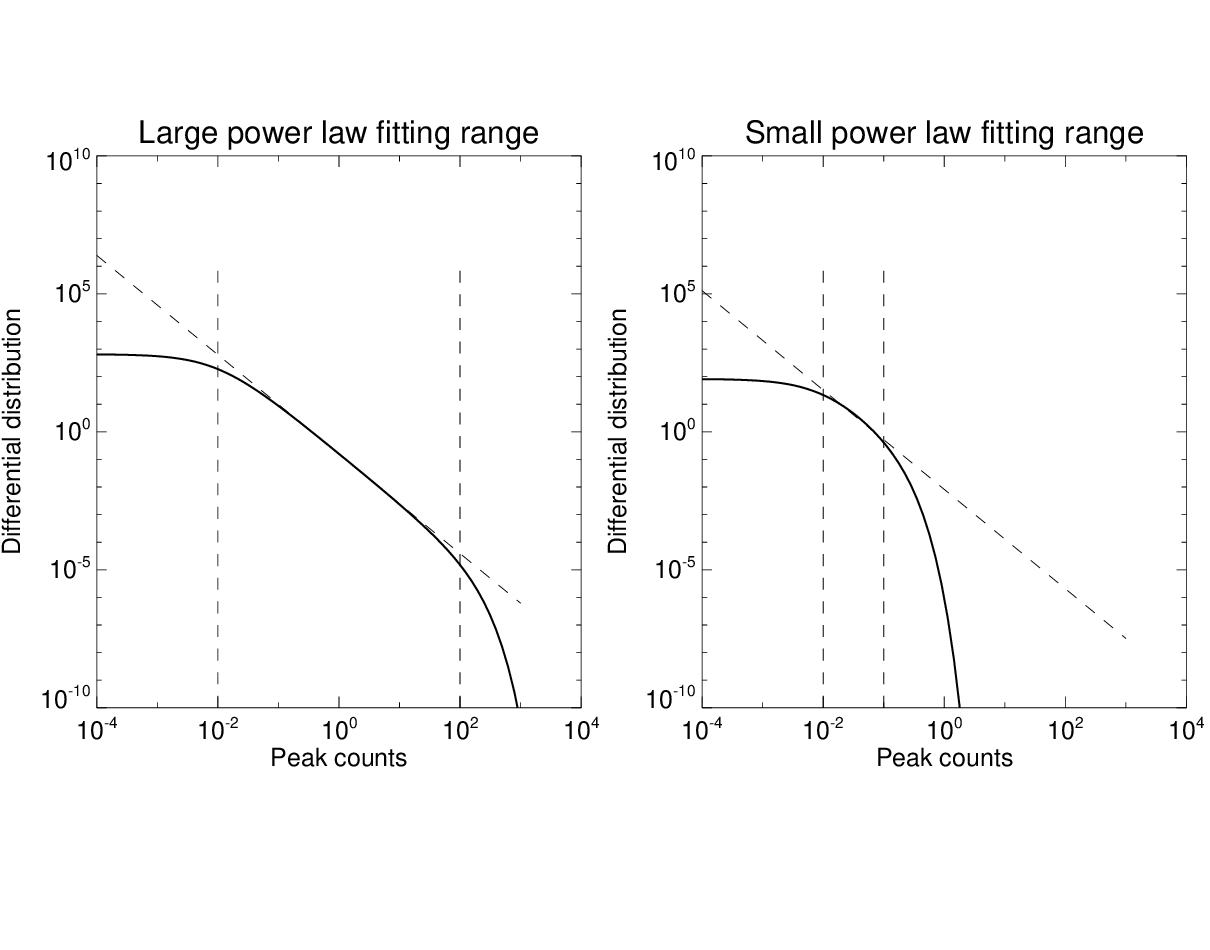}}
\caption{Power-law size distributions for large power law
fitting range (left-hand diagram) and for narrow fitting range
(right-hand diagram). Note the ambiguity of power law fitting
in the right-hand diagram.}
\end{figure}

\begin{figure}
\centerline{\includegraphics[width=1.0\textwidth]{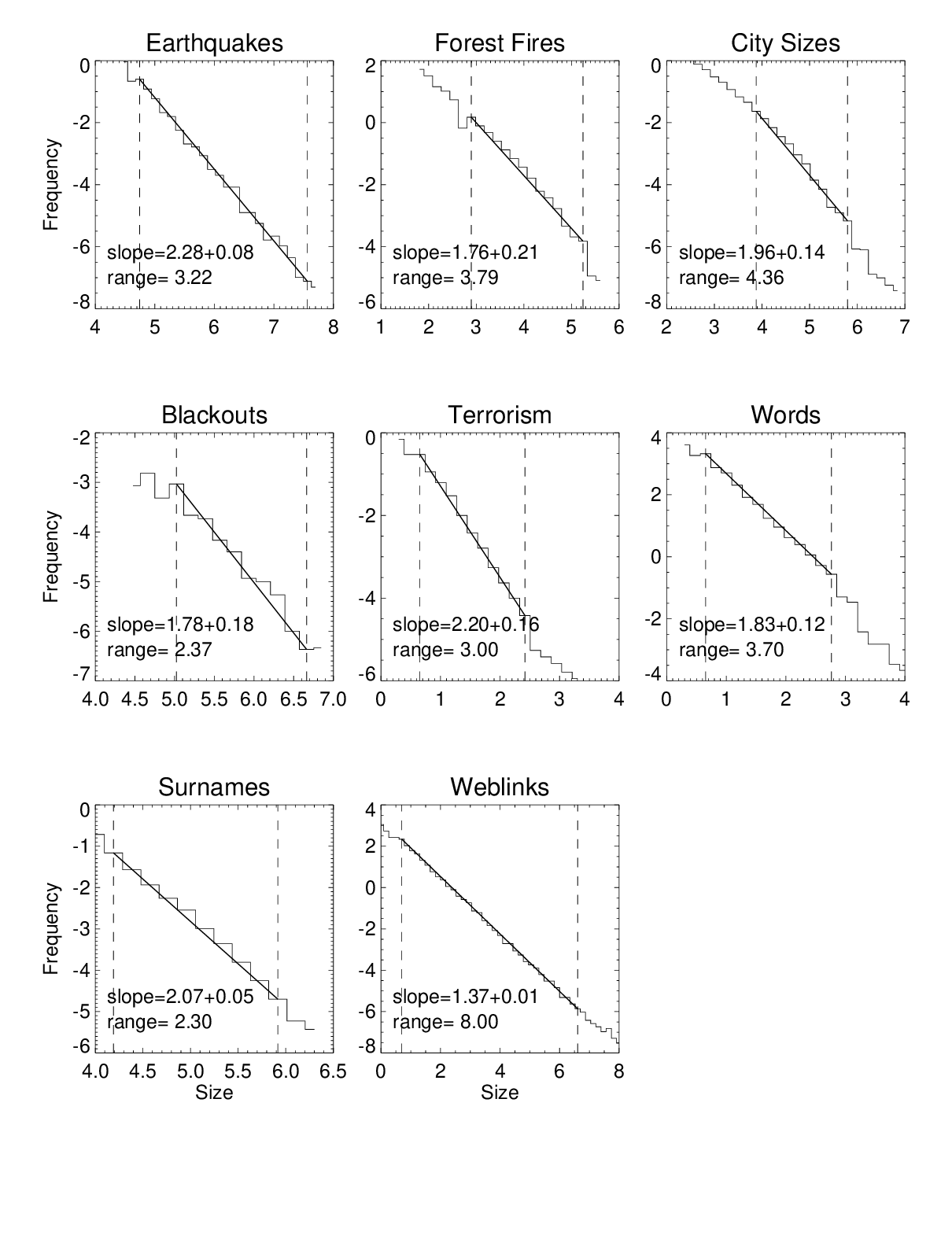}}
\caption{Power-law size distributions.}
\end{figure}

\end{document}